\documentstyle[eqsecnum,aps,prd,epsf]{revtex}
\renewcommand{\equiv}{:= }

\newcommand{\ads}{asymptotically de Sitter}

\newcommand{\AB}{A_B}
\newcommand{\AC}{A_C}
\newcommand\CC{{\Lambda}}
\newcommand{\paren}[1]{\left( #1 \right)}
\newcommand\unp{\tilde}
\newcommand\scri{{\cal I}}
\newcommand\R{{\bf R}}
\newcommand\N{{\bf N}}
\newcommand\interior{{\rm int}}

\newcommand{\gr}{{g_R}}

\newcommand{\EL}{E}

\newcommand{\RR}{{\cal R}}
\newcommand{\LL}{{\cal L}}
\newcommand{\DD}{{\cal D}}
\newcommand{\oo}[1]{{#1}_+}
\newcommand{\ii}[1]{{#1}_-}

\newtheorem{theorem}     {Theorem}
\newtheorem{lemma}       {Lemma}
\newtheorem{proposition} {Proposition}

\newenvironment{proof}{\noindent{\em Proof.}}%
{{\hspace*{1em}\hfill{$\Box$}}\\}
\newenvironment{corollary}{\noindent{\bf Corollary }\em}%
{\\}
{{\hspace*{1em} \hfill{$\Box$}}}
\newenvironment{remark}{\noindent{\bf Remark.}}{\\}

\begin{document}

\def\abstract#1{\begin{center}{\large Abstract}\end{center}
\par #1}

\title{Upper bound for entropy in asymptotically 
de Sitter space-time}
\author{Kengo Maeda%
  \thanks{Electronic address: maeda@th.phys.titech.ac.jp}, 
  Tatsuhiko Koike%
  \thanks{Electronic address: koike@rk.phys.keio.ac.jp},
  Makoto Narita%
  \thanks{Electronic address: narita@rikkyo.ac.jp},
  and Akihiro Ishibashi%
  \thanks{Electronic address: akihiro@th.phys.titech.ac.jp}}
\address{${}^{\mbox{\rm *\S}}$%
  Department of Physics, Tokyo institute of Technology, 
  Oh-Okayama, Meguro, Tokyo 152, Japan}
\address{${}^{\dag}$%
  Department of Physics, Keio University, Hiyoshi,
  Kohoku, Yokohama 223, Japan}
\address{${}^{\ddag}$%
  Department of Physics, Rikkyo University,
  Nishi-ikebukuro, Toshima, Tokyo 171, Japan}
\maketitle
\abstract{We investigate nature of asymptotically de Sitter space-times
containing a black hole. We show that if the matter fields satisfy 
the dominant energy condition and the cosmic censorship holds in the 
considering space-time, the area of the cosmological event horizon 
for an observer approaching a future timelike infinity 
does not decrease, i.e. the second law is satisfied. 
We also show under the same conditions that
the total area of the black hole and the cosmological event horizon,
a quarter of which is the total Bekenstein-Hawking entropy, 
is less than $12\pi/\Lambda$, where $\Lambda$ is a cosmological constant.
Physical implications are also discussed.}

\bigskip


\section{Introduction}
\label{introduction}
There has been interest in space-times with a positive cosmological
constant $\Lambda$. 
Recent cosmological observations suggest the existence of 
$\Lambda$ in our universe~\cite{lambda}.
Also, it is widely believed that the inflation took place in the 
early stage of our universe, where the vacuum energy of a scalar 
field (inflaton) plays a roll of $\Lambda$.
Most regions in such a space-time are expected to expand as in de
Sitter space-time. 
Some regions, however, will gravitationally collapse to form black
holes, if the inhomogeneity of initial matter distribution
is large.  Then there will be observers who have
two types of event horizons,
a black hole event horizon (BEH) 
and a cosmological event horizon (CEH), 
as the observers approaching the future timelike infinity
in Schwarzschild-de Sitter space-time.
Throughout this paper we shall focus event horizons for such
observers. 

Gibbons and Hawking~\cite{GH} studied thermodynamic property~\cite{H} 
of the event horizons in asymptotically de Sitter space-times.
In particular, they found that an observer 
feels a thermal radiation coming from the CEH 
and that the entropy $S_C$ of the CEH is equal to one quarter of 
its area as in the case of a BEH. 
Thus, the areas of the event horizons can be interpreted as the 
entropies, or lack of information of the observer
about the regions which he cannot see.
    
In classical general relativity, there have been a number of studies on 
the nature of BEH in the asymptotically de Sitter space-time. 
Hayward, Shiromizu and Nakao~\cite{HSN} 
and Shiromizu, Nakao, Kodama and Maeda~\cite{SNKM} 
showed that the area of a BEH in the
asymptotically de Sitter space-time cannot decrease and has an upper
bound $4\pi/\Lambda$ if the weak cosmic censorship (WCC)~\cite{P}
holds. It means that black holes cannot collide each other if the
total area of them exceeds the upper bound.

Davies~\cite{D} investigated a CEH in Robertson-Walker
models with $\Lambda$ and a perfect fluid satisfying the dominant
energy condition 
and showed that the area of the cosmological 
horizon cannot decrease.
From this result, one may expect that in generic 
asymptotically de Sitter space-times the area of a CEH cannot decrease 
as in the case of a BEH.

Boucher, Gibbons and Horowitz~\cite{BG} showed that the area of the CEH
is bounded from the above by $12\pi/\Lambda$ on a
regular time-symmetric hypersurface. 
Shiromizu, Nakao, Kodama and Maeda~\cite{SNKM} also obtained the 
same conclusion on a maximal hypersurface.
However, one cannot say that the same conclusion
holds for CEHs in a general non-stationary asymptotically de
Sitter space-time,
because it is highly nontrivial whether a foliation by such
hypersurfaces exists and covers the relevant portion of the space-time.

\begin{figure}[htbp]
 \centerline{\epsfxsize=8.0cm \epsfbox{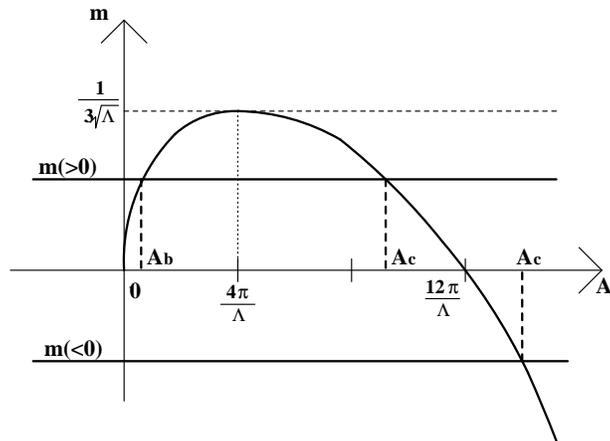}}
      \caption{
The mass parameter $m$ of a Schwarzschild-de Sitter solution 
for a fixed $\Lambda$ is related to the area $A$ of event horizons 
as $m =(A/16\pi)^{1/2}(1-\Lambda A/12\pi)$. 
$A_b$, $A_c$ are the areas of a BEH and a CEH, respectively.
}
          \protect
\label{Up-eps} 
\end{figure}

The WCC is assumed in the proof of the above results as well as in the 
case of a BEH.
An example of Schwarzschild-de Sitter space-time shows
significance of this assumption, and also 
suggests a close relation among the area of the CEH,
the WCC and positivity of the {\em gravitational energy} (mass).
Fig.~\ref{Up-eps} shows the mass parameter $m$ as a function of the 
area $A$ of the event horizon and 
Figs.~2(a) and 2(b) shows the Penrose diagrams for
the cases of $m>0$ and $m<0$, respectively.
One easily finds that if the WCC holds ($m>0$) the area ${\AC}$ of the
CEH is bounded from the above by $12\pi/\Lambda$.  
Indeed, one finds that the {\em total}\/ area of the BEH and the CEH
has an upper bound $12\pi/\Lambda$.  
On the other hand, if the WCC is violated ($m<0$) ${\AC}$ is not
bounded.  

In this paper, we show the area theorem that the area of the CEH in 
an asymptotically de Sitter space-time containing a black hole cannot
decrease so that the second law of thermodynamics is satisfied, 
and the total area of BEH and CEH is less than $12\pi/\Lambda$, 
hence total Bekenstein-Hawking entropy is less than
$3\pi/\Lambda$, if the space-time satisfies the WCC and the energy
conditions. 
To this end, we define a quasi-local energy in a space-time with
$\Lambda$ and its monotonicity and positivity. 
Very roughly speaking, our analysis is a generalization of the argument
of the previous paragraph to general asymptotically de Sitter
space-times which are neither stationary nor spherically symmetric. 

We follow the notation of Ref.~\cite{HE} and use the units 
$c=G=\hbar=k_B=1$.

\section{Asymptotically de Sitter space-time and the area law for
a cosmological event horizon}
\label{area-law}
In this section we shall show the area theorem 
(Theorem~\ref{th-area-law-CEH}) for a CEH in an asymptotically de Sitter
space-time. 

As a precise definition of an asymptotically de Sitter space-time
satisfying the WCC,  
we assume space-time $(M,g)$ to be
{\it strongly asymptotically predictable from a partial 
Cauchy surface $\Sigma$ and de Sitter in the future}~\cite{SNKM}, 
and just call it {\em \ads}.
In what follows, causal relationships are considered in a larger 
manifold $(\unp M,\unp g)$ in which $(M,g)$ is conformally embedded.
Note that the future conformal infinity $\scri^+$ of $M$ 
is a spacelike hypersurface in $\unp M$~\cite{PR}.

We shall consider {\ads} space-times containing a black hole 
and an observer whose world line $\lambda$ has a future endpoint 
at the ``future timelike infinity.'' 
Then $\dot J^-(\lambda)$ consists of two components, the BEH and the CEH
for the observer~\cite{GH}. 
As the BEH can be defined by $\dot J^-(\scri^+)$, the CEH can be also 
defined in terms of $\scri^+$. 
Namely, we define the {\em cosmological event horizon (CEH)}\/ to be the 
past Cauchy horizon $H^-(\scri^+)$ of the future infinity. 

In general, the topology of $\scri^+$ is not determined. 
However, it seems reasonable to suppose
that $\scri^+$ is diffeomorphic to ${\rm S}^2\times (0,1)$ if the
topology of the BEH is ${\rm S^2}$. 
In analogy of weakly asymptotically simple and empty, 
and future asymptotically predictable space-time 
(see Prop. 9.2.3 of Ref.~\cite{HE}), 
we also assume that there is a continuous onto map
$\alpha: (0,\infty)\times\Sigma\rightarrow D^{+}(\Sigma)-\Sigma\,$    
satisfying the following.
(1) 
For each $t\in (0,\infty)$, 
$\alpha_t:=\alpha(t,\cdot)$ and 
restriction of 
$\alpha$ on $(0,t)\times\alpha_t^{-1}(\Sigma_{t}-\scri^+)$ are
homeomorphisms, 
where $\Sigma_{t}\equiv \alpha(\{t\} \times \Sigma)$;
(2) For each $t\in (0,\infty)$, 
$\Sigma_{t}$ is a Cauchy surface for $D(\Sigma)$ 
such that 
(a) $\Sigma_{t_2}-\scri^+\subset I^{+}(\Sigma_{t_1}-\scri^+)$
when $t_2>t_1$,  
and 
(b) the edge of $\Sigma_{t}-\scri^+$ in $\unp{M}$ is a spacelike
two-sphere in $\scri^+$. 
We define 
$W_t \equiv \Sigma_{t}\cap\scri^+$.
We have 
$  W_{t_1}\subset W_{t_2}$ for $t_2>t_1$ and
$\bigcup_{t\in(0,\infty)} W_t=\scri^+$.

We also present a lemma about the topology of a CEH. 
\begin{lemma}
  (Each component of) any sufficiently nice 
  cut of the cosmological event horizon $H^-(\scri^+)$ is a
  topological two-sphere. 
\end{lemma}
\begin{proof}
  Since ${D}{}^-(\scri^+)\cap M$ is a future set in $M$, 
  its boundary in $M$, i.e., the 
  CEH, must be a $C^{1-}$ embedded submanifold of 
  $M$ (see Prop. 6.3.1 of Ref.~\cite{HE}).  
  Moreover, $\interior D^-(\scri^+)$ is simply connected
  because it is homeomorphic $\scri^+\times\R$ and $\scri^+$ is simply 
  connected. Thus the conclusion follows.
\end{proof}


We use the following lemma, which is shown in Ref.~\cite{HE}, to
prove Lemma~\ref{lem-D(Wn)}.
\begin{lemma}
  \label{lem-J-cpt}
  Let $\Sigma$ be a partial Cauchy surface. 
  For any $p\in D^-(\Sigma)$, $J^+(p)\cap D^-(\Sigma)$ is compact.
 \hfill$\Box$
\end{lemma}
\begin{lemma}
  \label{lem-D(Wn)}
  $D^-(\scri^+) = \bigcup_{t\in(0,\infty)} D^-(W_t)$.
\end{lemma}
\begin{proof}
  Let us define a continuous function 
  $\scri^+\ni p\mapsto t\in (0,\infty)$ defined by $p\in
  \mbox{edge}(W_t)$. 
  Because Lemma \ref{lem-J-cpt} implies that 
  for any $p\in D^-(\scri^+)$,
  $J^+(p)\cap \scri^+$ is compact in $\unp M$, 
  there exists a maximum value for the function above. 
  So there is a $t\in (0,\infty)$ such that 
  $W_t\supseteq J^+(p)\cap \scri^+$ and hence $p\in D^-(W_t)$.
  Thus we have $D^-(\scri^+)\subseteq\bigcup_{n\in\N} D^-(W_n)
  \subseteq\bigcup_{t\in(0,\infty)} D^-(W_t)$.
  It follows from 
  $D^-(\scri^+)\supseteq D^-(W_t)$ for each $t\in(0,\infty)$
  that $D^-(\scri^+)\supseteq \bigcup_{t\in(0,\infty)}D^-(W_t)$.
\end{proof}

In the next step we will prove Lemma~\ref{lem-seq} by using the
following Limit Curve Lemma~\cite{B}.
\begin{lemma}[Limit Curve Lemma]
  \label{lem-lim-crv}
  Let $\gamma_{n}:$  
  $(-\infty,\infty)\rightarrow M$ be a sequence of
  inextendible non-spacelike curves 
  (parametrized by arc length in $\gr$ which is a complete Riemannian metric).
  Suppose that $p\in M$ is an accumulation point of the sequence 
  $\{\gamma_{n}(0)\}$.
  Then there exist an 
  inextendible non-spacelike curve $\gamma$ such that
  $\gamma(0)=p$ and subsequence $\{\gamma_{m}\}$
  which converges to $\gamma$ uniformly
  (with respect to $\gr$) on compact subsets of ${\bf R}$.
\end{lemma}
\begin{lemma}
  \label{lem-seq}
  For any generator $\lambda$ of $H^-(\scri^+)$,
  parametrized with respect to $\gr$-arc length,
  there exists a sequence $\{\lambda_n\}$ of null geodesics
  in $D^-(\scri^+)$,
  parametrized with respect to $\gr$-arc length,
  such that 
  (1) $\{\lambda_n\}$ converges uniformly to $\lambda$ 
  with respect to $h$ on compact subsets of $\R$, and  
  (2) each $\lambda_n$ generates an achronal set. 
\end{lemma}
\begin{proof}
  Let $p$ be a point of $\lambda$ which is not the endpoint.
  Any neighborhood $U$ of $p$ contains a point of 
  $D^-(\scri^+)$. 
  It follows from Lemma \ref{lem-D(Wn)} that there exist a $n_0\in\N$ 
  such that $U\cap D^-(W_n)\ne\emptyset$ hence 
  $U\cap H^-(W_n)\ne\emptyset$ for all $n\ge n_0$.
  Then one can construct a sequence $\{p_n\}$ such that
  $p_n\in H^-(W_{m_n})$ and $p_n\rightarrow p$, where 
  $(W_{m_n})$ is a subsequence of $\{W_n\}$.
  Letting $\lambda_n$ be the generator of $D^-(W_{m_n})$ through
  $p_n$, one has from Lemma~\ref{lem-lim-crv} that there exists a
  inextendible non-spacelike $C^0$-curve $\gamma$ through $p$ such
  that $\{\lambda_n\}$ converges to $\gamma$ uniformly on compact
  subsets of $\R$.   
  However, because $\{\lambda_n\}$ can have its accumulation points  
  only on $H^-(\scri^+)$, $\gamma$ must lie on 
  $H^-(\scri^+)$.
  Since $\gamma$ is a non-spacelike curve through $p$ and is lying on 
  $H^-(\scri^+)$, it must coincide with $\lambda$. 
\end{proof}

Finally we present the following area theorem of the CEH.
\begin{theorem}[Area law for a CEH]
  \label{th-area-law-CEH}
  In an {\ads} space-time with a piecewise smooth CEH satisfying the
  weak energy condition, 
  $A(H^-(\scri^+)\cap\Sigma_{t_2}) \ge A(H^-(\scri^+)\cap\Sigma_{t_1})$
  for $t_2>t_1$,
  where $A({\cal S})$ denotes the area of a two-surface ${\cal S}$. 
\end{theorem}
\begin{proof} 
  Piecewise smoothness of the CEH implies that 
  there are a finite number of 
  pairwise disjoint smooth submanifolds $U_i$'s such that the CEH
  is $\bigcup_i \overline U_i$. 
  It suffices to show that the expansion $\theta\ge0$ 
  on each $p\in\mbox{\rm int}\,U_i$ because 
  each $U_i$ is foliated by future inextendible null geodesic
  generators.
  For any point $p\in\mbox{\rm int}\,U_i$ for some $i$  
  there is an open set $V\ni p$ 
  diffeomorphic to ${\cal S}\times\R$
  where ${\cal S}$ is a locally spacelike two-surface containing
  $p$ with compact closure. 
  By Lemma~\ref{lem-lim-crv} and compactness of 
  $\overline{\cal S}$ there is a sequence of diffeomorphisms 
  $\phi_n:V\to V_n\subset H^-(W_n)$ 
  such that 
  (1) each $\phi_n({\cal S})$ is spacelike, 
  (2) each $\phi_n$ preserves the foliations by null geodesic
  generators,
  and
  (3) $\phi_n(V)$ converges uniformly to $V$ on compact 
  subsets of ${\cal S}\times\R$.
  Suppose the expansion $\theta$ of future-directed null geodesic
  generators of the 
  CEH was negative at $p$. 
  Then by the continuity of $\theta$ 
  there would be some $n$ such that 
  the expansion $\theta_n$ of generators of $V_n$ was negative at
  $\phi_n(p)$. 
  From the weak energy condition 
  the generator from $\phi_n(p)$, since it is future complete, 
  would have a conjugate point of $\phi_n({\cal S})$~(see Prop. 4.4.6
  of Ref.~\cite{HE}).
  This contradicts achronality of $H^-(W_n)$.
\end{proof}

\begin{corollary}
  If the assumptions of Theorem \ref{th-area-law-CEH} hold
  and 
  every future incomplete null geodesics terminates in a strong 
  curvature singularity of Kr\'{o}lak~\cite{K}, 
  then every generator of the CEH is future complete.
\end{corollary}

\begin{proof}
From the proof of Theorem~\ref{th-area-law-CEH}, 
the expansion of each null geodesic
generator cannot be negative. This contradicts the condition of 
the strong curvature singularity.  
\end{proof}

\section{Quasi-local energy in space-times with $\Lambda$}
\label{QLE}

We define a quasi-local energy $\EL({\cal S})$ in a
space-time with $\Lambda$ and examine its monotonicity and positivity, 
which we will use to show the existence of 
an upper bound for entropy (Theorem~\ref{th-up-EH}) in
Sec.~\ref{upper-bound-theorem}. 

Let us introduce Hayward's double null formalism~\cite{H1}, 
namely, smooth foliations of null three-hypersurfaces
labeled by $\xi_\pm$ such that 
each intersection of two hypersurfaces of constant
$\xi_\pm$ is a closed spacelike two-surface.
We have the evolution vector $u_\pm=\partial/\partial\xi_\pm$, 
the normal one-forms $n_{\pm}=-d\xi_{\pm}$, 
the metric $h=g+e^{-f}(n_+ n_- + n_- n_+)$ induced on the two-surface,
the projection $\perp$ on the two-surface,
the shift vectors $r_\pm=\perp\! u_\pm$, and 
the null normal vectors $l_\pm=u_\pm-r_\pm$.
The expansions $\theta_{\pm}$, 
the shears ${\sigma}^{\pm}$
and the twist $\omega$ on a two-surface are defined as      
\begin{eqnarray}
\theta_{\pm} &=& \frac{1}{2} h^{-1}:{\cal L}_{\pm}h,\\  
\sigma^{\pm} &=& {\cal L}_{{\pm}}h-\theta_{\pm}h,\\
\omega &=& \frac{1}{2}e^{f}h\cdot[l_{+},l_{-}],
\end{eqnarray}
where ${\cal L}_{{\pm}}$ represents the Lie derivatives along the vector
fields $l_{\pm}$, and a dot and a colon denote single and double
contraction, respectively.
The quasi-local energy is defined in each embedded spatial two-surface
${\cal S}$ 
as
\begin{eqnarray}
  \label{eq-Hawking-Lambda}
  \EL({\cal S})&:=& \frac{1}{8\pi}\sqrt{\frac{A}{16\pi}}\int_{\cal S}
  \mu\paren{\RR + e^f\oo\theta\ii\theta-\frac{2\Lambda}{3}},
\end{eqnarray}
where $A$, $\RR$, and $\mu$ represent the total area of ${\cal S}$,
the Ricci scalar on ${\cal S}$, 
and the area 2-form on ${\cal S}$, respectively.   
This is the Hawking energy 
with the last term added in the integrand.
Physically, $\EL({\cal S})$ is the gravitational energy subtracted by the
energy due to the cosmological constant $\Lambda$,
so that it is considered as the energy of the matter fields.
In Schwarzschild-de Sitter space-time
$\EL({\cal S})$ coincides with the mass parameter $m$.
In spherically symmetric space-times with dust
$\EL({\cal S})$ coincides with the mass function~\cite{nakao}.
In space-times without $\Lambda$ our quasi-local
energy $\EL({\cal S})$ reduces to the Hawking energy.

The Einstein equations are given by
\begin{eqnarray} 
\label{eq-Ray}
e^{-f}{\cal L}_{\pm}(e^f\theta_{\pm})+
\frac{1}{2}{\theta_{\pm}}^2 +\frac{1}{4} 
{\|\sigma_{\pm}\|}^2 &=& -8\pi\phi_{\pm},
\end{eqnarray}
\begin{eqnarray}
\label{eq-cross}
&&{\cal L}_{\pm}\theta_{\mp}+\theta_{+}\theta_{-}+
e^{-f}\left[\frac{1}{2}\RR-{\left|\frac{1}{2}\DD f\pm\omega\right|}^2+
  \DD\cdot\left(\frac{1}{2}\DD f\pm\omega\right)\right] \nonumber \\
&&\qquad\qquad = 8\pi\rho+e^{-f}\Lambda,
\end{eqnarray}
where $\phi_{\pm}=T(l_{\pm},l_{\pm})$ and $\rho=T(l_{+},l_{-})$ for 
the energy tensor $T$, 
and $\DD$ is the covariant derivative with respect to $h$.

Let us examine the monotonicity of $\EL({\cal S})$ 
on an outgoing null hypersurface $\xi_{-}=$ {\em constant}\/ 
(the monotonicity on an ingoing null hypersurface $\xi_{+}=$ {\em constant}\/
or on a spacelike hypersurface can be argued similarly.) 
The derivative of the energy $\EL({\cal S})$ along the outgoing
direction $l_{+}$  
is  
\begin{eqnarray}
\label{eq-mono}
 8\pi{\cal L}_{+}\EL &=& \sqrt{\frac{A}{16\pi}}\Biggl[ \frac{1}{2A}
 \int_{\cal S}\mu\theta_{+}\int_{\cal S}\mu
 \left( \RR+e^f\theta_{+}\theta_{-} \right) \nonumber \\     
 & & {} - \int_{\cal S}\mu\theta_{-}\left( \frac{1}{4}
{\|\sigma_{+}\|}^2 + 8\pi\phi_{+} \right) \nonumber \\
 & & {} - \int_{\cal S}\mu\theta_{+} \Bigl( \frac{1}{2}\RR + \frac{1}{2}e^f
 \theta_{+}\theta_{-} + \DD\cdot \bigl( \frac{1}{2}\DD f 
 + \omega \bigr) \nonumber \\
 & & {} - {\left| \frac{1}{2}\DD 
 f + \omega \right|}^2  
 - 8\pi  e^f\rho \Bigr) \Biggr]. 
\end{eqnarray}  
We assume that the matter fields satisfy the dominant energy condition, 
$\phi_{+}\ge 0$ and $\rho\ge 0$,  
and take a foliation of the hypersurface $\xi_-=$ {\em constant}\/
by spatial two-surfaces ${\cal S}$.
The energy $\EL({\cal S})$ is non-decreasing in the outgoing 
null direction($\theta_{+}\ge 0$, $\theta_{-}\le 0$), 
${\cal L}_{+}\EL\ge 0$, if 
\begin{eqnarray}
{\langle \theta_{+} \rangle}{\langle F \rangle}\ge
\langle \theta_{+}F \rangle
\label{uef}
\end{eqnarray}
on each ${\cal S}$, where 
\begin{eqnarray}
&&F\equiv \RR + e^f\theta_{+}\theta_{-} + 2\DD\cdot 
(\frac{1}{2}\DD f + \omega),\\
&&\langle \cdot \rangle\equiv 
\frac{\int_{\cal S}\mu \;\cdot\;\;}{\int_{\cal S}\mu}.
\end{eqnarray}
We remark that each term of $F$ except the third term  
is invariant under rescaling of the outgoing null normal $l_+$.  
An example of the foliations satisfying Eq.~(\ref{uef}) is one 
with $F=$ {\em constant}, 
which we can take by the rescaling of $l_+$.
Another example is the 
uniformly expanding foliation~\cite{H2}.

\section{Upper bound for the area}
\label{upper-bound} 

In this section we will show that the total area of the BEH and the
CEH is bounded in {\ads} space-times (Theorem~\ref{th-up-EH}).

We define the apparent horizons according to 
Hayward~\cite{H3}. 
A {\em marginal surface}\/ is a
spatial two-surface ${\cal S}$ on which ${\theta_{+}}=0$ or 
${\theta_{-}}=0$.
A {\em black hole apparent horizon (BAH)}\/ is 
the closure $\overline{T_B}$ of a hypersurface $T_B$ 
foliated by marginal surfaces 
on which
${\theta_{+}}=0$,
${\theta_{-}}<0$ and ${\cal L}_{-}{\theta_{+}}<0$. 
A {\em cosmological apparent horizon (CAH)}\/ is $\overline{T_C}$ 
foliated by marginal surfaces 
on which ${\theta_{-}}=0$, 
${\theta_{+}}>0$ and ${\cal L}_{+}{\theta_{-}}>0$.
Here the coordinates $\xi_\pm$ are taken so that they are constant on
each of the above spatial two-surfaces.
   
Hayward, Shiromizu and Nakao~\cite{HSN} showed that the area of a BAH
has an upper bound $4\pi/\Lambda$. 
They also showed that the area of a BEH
is less than $4\pi/\Lambda$ by implicitly
assuming the existence of the limit two-surface ${\cal S}$ of the 
BEH, though its physical meaning is not clear     
(see Appendix). 
Instead, 
one can reach the same conclusion 
under a physically reasonable condition,
strongly future asymptotically predictability (or WCC)
in an ``extended'' sense~\cite{I}. 
It states 
that singularities are hidden inside not only a BEH but also a BAH. 
More precisely, the closure of the domain of dependence of a 
partial Cauchy surface contains not only $\scri^+$ and the BEH but
also the outermost part of the BAH, i.e., 
(i) there exist $t>0$ and a subset $T_B'$ of ${T_B}$, foliated by
marginal surfaces, such that 
$H^{-}(\overline{T_B'})\cap J^{+}(\Sigma_t)\supseteq
\dot J^{-}(\scri^{+})\cap J^{+}(\Sigma_t)$ and 
$[I^{-}(\overline{T_B'})\cap I^{+}(\Sigma_t)]\subseteq D^{+}(\Sigma_t)$. 
We have the following proposition, 
whose proof we give in Appendix. 
\begin{proposition}
\label{th-up-BEH}
In an {\ads} space-time satisfying condition (i) above and the weak
energy condition, 
the area of a black hole event horizon (BEH) is
less than $4\pi/\Lambda$. 
\end{proposition}

\label{upper-bound-theorem}
Now we will show that the total area of BEH and CEH 
has an upper bound $12\pi/\Lambda$ by making use of
Proposition~\ref{th-up-BEH}. 
We require the following conditions.
(ii) There exists $t_0\ge0$ such that the cross section of 
  $\dot J^-(\scri^+)\cap \Sigma_{t}(t\ge t_0)$ 
  is smooth one connected component and the topology is ${\rm S}^2$;
(iii) There exists a marginal surface ${\cal S}_t$ with $\theta_{-}=0$
whose topology is ${\rm S}^2$ in each $\Sigma_{t}(t\ge t_0)$ and surrounds 
  $\dot J^-(\scri^+)\cap \Sigma_{t}$; 
(iv) $I^{-}(T_C)\cap J^{+}(\Sigma_{t_0})=(I^{-}({\cal I}^{+})-
 \overline{D^{-}({\cal I}^+)})\cap J^{+}(\Sigma_{t_0})$.
(v) any null geodesic generator of BEH is future complete. 
(vi) Matter fields satisfy the dominant energy condition. 
[This implies that matter field satisfies 
the the weak energy condition~(see e.g., Ref.~\cite{HE}).] 
(vii) There exists a foliation satisfying Eq. (\ref{uef}) on each
outgoing null hypersurface $\dot{J}^{-}({\cal S}_t)$ inside the CAH. 
Condition (iv) is similar to condition (i) above.
\begin{lemma}
\label{lem-pos-qlm}
For an arbitrary small positive value, $\epsilon_1$,
there is an acausal hypersurface $\Sigma_{t_1}(t_1>t_0)$  
such that for any closed spacelike two-surface ${{\cal S}_B}$ of 
${\dot{J}}^{-}({\cal I}^{+})\cap J^{+}(\Sigma_{t_1})$  
the quasi-local energy $\EL({{\cal S}_B})$ satisfies 
\begin{eqnarray} 
\label{es}
\EL({{\cal S}_B}) &\ge& 
\frac{1}{8\pi}\sqrt{\frac{A({{\cal S}_B})}{16\pi}}
\left(8\pi-\epsilon_1-\frac{2\Lambda}{3}A({{\cal S}_B})\right)>0.
\end{eqnarray}
\end{lemma}
\begin{proof}
Consider each null geodesic generator $l_{+}$ of the BEH.
By condition (v), (vi) and the Raychaudhuri equation~(\ref{eq-Ray}), 
$\lim_{\xi \to \infty} \theta_{+} =0$ is satisfied, where 
$\xi$ is an affine parameter of $l_{+}$.  
$\lim_{\xi \to \infty} \int_{\cal S}\mu e^f\theta_{+}\theta_{-} =0$ 
is also satisfied because the area of a BEH has an upper bound. 
Therefore there is a $\Sigma_{t_1}$ such that for any closed spacelike 
two-surface ${{\cal S}_B}$ of 
${\dot{J}}^{-}({\cal I}^{+})\cap J^{+}(\Sigma_{t_1})$,  
$\int_{\cal S}\mu e^f\theta_{+}\theta_{-}$ is larger than  
$ - \epsilon_1$, where $\epsilon_1$ is an arbitrary small positive
value. From Eq.~(\ref{eq-Hawking-Lambda}) and Prop.~\ref{th-up-BEH}
one can get the desired result by using  
the Gauss--Bonnet theorem and condition (ii).
\end{proof}
\begin{lemma}
\label{lem-marg}
$J^{-}(T_C)\cap J^{+}(\Sigma_{t_0})=
(\bigcup_{t}J^{-}({\cal S}_t))\cap J^{+}(\Sigma_{t_0})$.
\end{lemma}
\begin{proof}
For any point $p\in J^{-}(T_C)\cap J^{+}(\Sigma_{t_0})$
there is a point $q\in J^+(p)\cap T_C$.  Then
there is ${\cal S}_t\ni q$ so that $p \in J^{-}({\cal S}_t)$. 
\end{proof}
\begin{theorem}
\label{th-up-EH}
If an {\ads} space-time satisfies the conditions (i)--(vii) above, 
${\AB}\equiv\lim_{\xi_+\rightarrow \infty}A({{\cal S}_B})$ and 
${\AC}\equiv\lim_{\xi_-\rightarrow \infty}A({{\cal S}_C})$ satisfy 
\begin{eqnarray} 
\label{eq-up-EH}
{\AB}+{\AC}+\sqrt{{\AB}{\AC}}\le\frac{12\pi}{\Lambda}. 
\end{eqnarray}
\end{theorem}
\begin{remark}
  In particular, the area $A_C$ of the CEH is less than
  $12\pi/\Lambda$. 
\end{remark}
\begin{proof}
For any closed spacelike two-surface ${{\cal S}_C}$ of 
$H^{-}({\cal I}^{+})\cap D^+(\Sigma)$ 
there exists a partial Cauchy surface $\Sigma_{{{\cal S}_C}}$ 
containing ${{\cal S}_C}$.
Consider a sequence of marginal surfaces 
${{\cal S}_n}$ ($n\in {\bf N}$) defined above and
define $N^+_n$ and $N^-_n$ 
as the null hypersurfaces generated by the future-directed outgoing and
ingoing null geodesic generators of $\dot{J}^-({\cal S}_n)$, respectively. 
Denote the spacelike two-surface 
$N^-_n\cap\Sigma_{{{\cal S}_C}}$
by ${\cal K}_n$. From the condition (iv) and Lemma~\ref{lem-marg}
it follows that 
$ \lim_{n \rightarrow \infty} A({{\cal K}_n}) = A({{\cal S}_C})$.
The expansion $\theta_{-}$ of $l_{-}$ is non-negative 
in the future direction between ${\cal S}_n$ and ${\cal K}_n$ 
because $\LL_-(e^f\theta_{-})\le0$ there, 
as implied by the Raychaudhuri equation~(\ref{eq-Ray}) of $l_{-}$ and
condition (vi), and by
$\theta_{-}=0$ on each ${\cal S}_n$. 
Thus, as in the proof of Prop.~\ref{th-up-BEH}, 
there exists $n_1$ for an arbitrary small positive
value $\epsilon_2$ such that for all $n>n_1$ 
\begin{eqnarray}
A({{\cal S}_C})-\epsilon_2\le A({{\cal S}_n})
\end{eqnarray} 
is satisfied.  

Consider 
outgoing null hypersurfaces 
$\hat N^+_n \equiv N^+_n\cap J^{+}(\Sigma_{t_1}) 
\cap J^{+}(\Sigma_{{\cal S}_C})$. 
From the condition (iv) and Lemma~\ref{lem-marg}, 
for any neighbourhood $U$ of 
${\dot{J}}^{-}({\cal I}^+)\cap J^{+}(\Sigma_{t_1})
\cap J^{+}(\Sigma_{{\cal S}_C})$, 
there is $ n_2 > n_1$ such that for $n>n_2$ each $\hat N^+_n$ 
intersect $U$. 
For $n>n_2$, take spacelike two-surfaces ${\cal Q}_n$ 
in $\hat N^+_n\cap U$. 
The sequence $\{{\cal Q}_n\}$ converges 
to a spacelike two-surface ${\cal S}_B$ of 
${\dot{J}}^{-}({\cal I}^{+})\cap J^{+}(\Sigma_{t_1})
\cap J^{+}(\Sigma_{{\cal S}_C})$. 
By the continuity of $\EL({\cal Q}_n)$, 
for an arbitrary small 
$\epsilon_3 > 0$ there is $n_3>n_2$ such that 
$\EL({\cal S}_B)-\epsilon_3\le\EL({\cal Q}_n)$ for each $n>n_3$. 
By condition (vii), the energy $\EL({\cal S})$  
is non-decreasing from ${\cal Q}_n$ to ${\cal S}_n$ on $\hat N^+_n$. 
Thus $\EL({\cal S}_B)-\epsilon_3\le\EL({\cal S}_n)$ for each $n>n_3$. 
By Lemma~\ref{lem-pos-qlm}
and Eq.(\ref{eq-Hawking-Lambda}) for ${{\cal S}_n}$,  
this inequality can be rewritten as 
\begin{eqnarray} 
A({{\cal S}_B})+A({{\cal S}_n})&\le &
\frac{12\pi}{\Lambda}-\sqrt{A({\cal S}_B)A({{\cal S}_n})}
\nonumber\\
&&+O(\epsilon_1)+O(\epsilon_3).
\end{eqnarray}
Since ${{\cal S}_C}$ is an arbitrary two-surface of $H^{-}({\cal I}^{+})$,
one gets the desired result by taking limit $\epsilon_1, \epsilon_3\to0$.
\end{proof}

\section{Conclusions and Discussion}
\label{conclusion}

We have shown in Theorem~\ref{th-area-law-CEH} that in an {\ads}
space-time the area $A({\cal S}_C)$ of the CEH is non-decreasing if
the WCC and the weak energy condition hold. 
This means that the area law of event horizons holds not only 
for a BEH but also for a CEH hence it also applies to the total 
area of event horizons(total Bekenstein-Hawking entropy, i.e. a 
quarter of the total area of the BEH and the CEH). 
Next we have shown in Theorem~\ref{th-up-EH} 
that the final values of the areas satisfy
${\AB}+{\AC}+\sqrt{{\AB}{\AC}}\le{12\pi}/{\Lambda}$. 
This means that the final values of entropies
$S_B:=A_B/4$ of the BEH and $S_C:=A_C/4$ of the CEH, 
satisfy 
\begin{equation}
{S_B}+{S_C}+\sqrt{{S_B}{S_C}}\le{3\pi}/{\Lambda}. 
\label{eq-desired}
\end{equation}
In particular, the total entropy is bounded from the above by
$3\pi/\CC$ in an {\ads} space-time.
We note that the inequality in Theorem~\ref{th-up-EH} is stronger 
than the previous result and conjecture which 
state that $A({\cal S}_B)\le 4\pi/\Lambda$ and 
$A({\cal S}_C)\le 12\pi/\Lambda$. 

As discussed in Ref.~\cite{GH}, a BEH is unstable against 
the Hawking radiation, while a CEH is stable.    
Physically this suggests that all asymptotically de Sitter space-times 
approach de Sitter space-time. 
This is consistent with the inequality~(\ref{eq-desired}) 
which states that for a fixed $\Lambda$ the total entropy attains its
maximum in de Sitter space-time, although the quantum effects 
were not taken into account in the derivation of the inequality. 
This curious correspondence suggests that the inequality 
is another law of EH thermodynamics in asymptotically de Sitter space-times. 

It is of interest to pursue connections 
of the present result with the cosmic no hair
conjecture~\cite{GH,CNH}. 
Here we consider a weaker version of the conjecture which states 
that a space-time with $\CC$ has a future asymptotic region
rather than is recollapsing, 
i.e., the space-time is future {\ads}.
Since the areas of the EHs have universal bound (i.e., are bounded by
numbers which depend only on $\CC$)
and the areas are expected to become larger
when matter falls into them, one expects that the amount of matter
which falls into the EHs have a universal bound. 
So, in collapse of an isolated object, 
if most of the matter falls into either the BEH or the CEH, 
i.e., if there will be no heavy shell-like ``star'' surrounding the
black hole, 
one can expect that the total initial energy of the matter should be
bounded by a number which only depends on $\CC$.
This may provide a criterion for the existence of the future
asymptotic region of the space-time, that is, a criterion for   
the validity of the cosmic no hair conjecture.

To solve the problems above, it is very important to know the
property of the total entropy $S_T$ of the universe, 
i.e. the sum of the entropy of the EHs and 
that of the matter between the EHs.
We conjecture that in {\ads} space-time 
$S_T$  is non-decreasing, i.e., the generalized
second law of thermodynamics holds, and also
$S_T$ is bounded.

\section*{Acknowledgments}
We express our special thanks to Professor H. Kodama for
critical comments and fruitful discussions at the early stage of the
work.  
We thank Dr. S. Hayward, Dr. T. Mishima, Dr. T. Okamura and
Dr. T. Shiromizu for 
discussions, and Professor A. Hosoya and Professor
H. Ishihara for encouragement.  
The work is supported in part by 
the Japan Society for Promotion of Science (K.M.) and 
the Ministry of Education, Science, Sports and Culture of Japan 
(T.K.).

\appendix
\section{Proof of Proposition 1}
In their proof of the theorem of the upper bound for the area of the 
BEH in Ref.~\cite{HSN} 
Hayward, Shiromizu and Nakao 
implicitly assumed that there is a limit 
two-surface of the BEH 
on which quantities such as ${\cal L}_{-}{\theta_{+}}$ are
continuous, i.e., independent of how one approaches the ``timelike
infinity'' $i^+$. 
However, this is physically not very well motivated and 
is highly nontrivial in general. 
Here, we shall drop the assumption above and prove a slightly
modified version of the theorem. 

{\it Proof of Proposition~\ref{th-up-BEH}.}
It is enough to show that the area of a BEH in $J^{+}(\Sigma)$
has an upper bound because the area does not decrease
in the future direction as shown in Ref.~\cite{SNKM}.  
Let us consider a sequence of marginal surfaces ${\cal S}_n$ 
with $\theta_{+}=0$ on the BAH $T_B$ 
and take a sequence of subsets $T_n$ of $T_B'$ 
such that $T_{n-1} \subset T_n$, $\mbox{edge}(T_n) = {\cal S}_n$,
$\bigcup_{n\in\N}T_n=T'_B$. 
Consider spacelike two-surfaces
${\cal T}\equiv\dot J^{-}({\cal I}^{+}) \cap \Sigma_{t}$
and ${\cal T}_n\equiv H^{-}(T_n)\cap \Sigma_t$
for some (sufficiently large) fixed $t$. 
We can observe $D^{-}(T_B')=\bigcup_{n\in\N}D^{-}(T_n)$, by replacing 
${\scri}^{+}$ in Lemma \ref{lem-D(Wn)} with $T_B'$.
This together with condition (i)
implies that the sequence 
${\cal T}_n$ converges to ${\cal T}$. 
The expansion $\theta_{+}$ of each null geodesic generator $l_{+}$ of 
$H^{-}(T_n)$ is non-negative 
in the future direction between ${\cal T}_n$ and ${\cal S}_n$
because $\LL_+(e^f\theta_{+})\le0$ there, 
as implied by the Raychaudhuri equation~(\ref{eq-Ray}) and
the weak energy condition, and by 
$\theta_{+}=0$ on each ${\cal S}_n$. 
Thus, 
\begin{eqnarray}        
 \label{eq-A-1}
A({\cal T}_n)\le A({{\cal S}_n})\le \frac{4\pi}{\Lambda},
\end{eqnarray}
where the second inequality is obtained by integrating
Eq. (\ref{eq-cross}) multiplied by $e^f$ on marginal surface 
${\cal S}_n$ and using the Gauss-Bonnet theorem~\cite{HSN}.
Since the sequence ${\cal T}_n$ 
converges to ${\cal T}$, for arbitrary small $\epsilon>0$
there exists a $n_0\in {\bf N}$ such that $A({{\cal T}_n})$ 
with $n>n_0$ satisfies
\begin{eqnarray}
 \label{eq-A-2}
 A({\cal T})-\epsilon\le A({{\cal T}_n}).
\end{eqnarray} 
From inequalities (\ref{eq-A-1}) and (\ref{eq-A-2}) we have
\begin{eqnarray}
 A({\cal T})-\epsilon\le \frac{4\pi}{\Lambda}.
\end{eqnarray} 
Since this holds for any $\epsilon$ we have
\begin{eqnarray}
A({\cal T})\le \frac{4\pi}{\CC}.
\end{eqnarray} 

\hfill$\Box$

\begin{figure}[htbp]
 \centerline{\epsfxsize=8.0cm \epsfbox{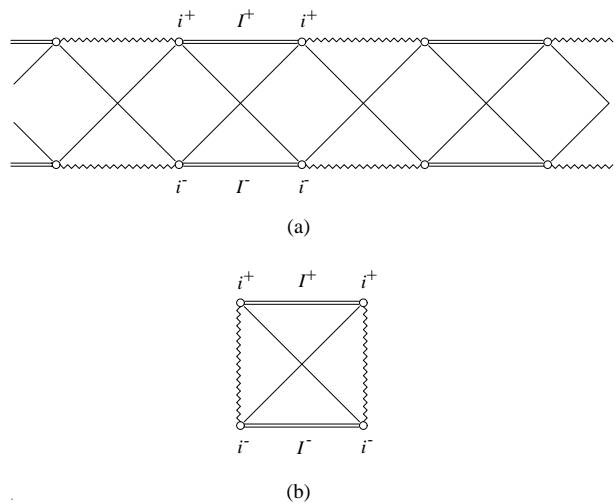}}
        \caption{Penrose diagrams of Schwarzschild-de Sitter space-times
with mass parameters (a)$m>0$, and (b)$m<0$, respectively.}
             \protect
\end{figure}


\begin{thebibliography}{99}


\bibitem{lambda}
M. Fukugita, F. Takahara, K. Yamashita and 
Y. Yoshii, Astrophys.  J.  (Letters)  {\bf 361} (1990)  L1, 
T. Totani, K. Sato, and Y. Yoshii, Astrophys.  J.  {\bf 460} (1996)  303, 
G. Efstathiou, W. J. Sutherland and S. J. Maddox, 
Nature {\bf 348} (1990)  705, 
J. A. Peacock and S. J. Dodds, Mon.  Not.  R.  Astron.  Soc. {\bf 267}
(1994)  1020.  

\bibitem{GH}G. W. Gibbons and S. W. Hawking, 
Phys. Rev. {\bf D15} (1977) 2738. 

\bibitem{H} J. Beckenstein,
Phys. Rev. {\bf D7} (1973) 2333, 
Phys. Rev. {\bf D9} (1974) 3292.  
S. W. Hawking, Comm. Math. Phys. {\bf 43} (1975) 199, 
Phys. Rev. {\bf D13} (1976) 191. 

\bibitem{HSN}S. A. Hayward,T. Shiromizu and K. Nakao, 
Phys. Rev. {\bf D49} (1994) 5080. 

\bibitem{SNKM}T. Shiromizu, K. Nakao, H. Kodama and K. Maeda, 
Phys. Rev. {\bf D47} (1993) R3099. 

\bibitem{P}R. Penrose, Riv.  Nuovo Cimento {\bf 1} (1969) 252. 

\bibitem{D}P. C. W. Davies,
Class. Quantum. Grav. {\bf 4} (1987) L225. 

\bibitem{BG}W. Boucher,G. W. Gibbons and G. Horowitz, 
Phys. Rev. {\bf D30} (1984) 2447. 

\bibitem{HE}S. W. Hawking and G. F. R. Ellis, 
{\em The large scale structure of space time} 
(Cambridge University Press, Cambridge, 1973) . 

\bibitem{PR}
{R.  Penrose and W.  Rindler},
{\em Spinors and Spacetime vol.  2}
(Cambridge University Press, Cambridge, 1986) . 

\bibitem{B}J. K. Beem, P. E. Ehrlich and K. L. Easley, 
{\em Global Lorentzian Geometry, 2nd ed. } 
(Marcel Dekker, New York, 1996) . 

\bibitem{K}A. Kr\'{o}lak,
J. Math. Phys. {\bf 28} (1987) 2685. 

\bibitem{H1}S. A. Hayward,
Phys. Rev. {\bf D49} (1994) 831. 

\bibitem{nakao}K. Nakao, preprint gr-qc/9507022. 

\bibitem{H2}S. A. Hayward,
Class. Quantum Grav. {\bf 11} (1994) 3037. 


\bibitem{H3}S. A. Hayward, Phys. Rev. {\bf D49} (1994) 6467. 


\bibitem{I}W. Israel,
Phys. Rev. Lett. {\bf 57} (1986) 397. 

\bibitem{CNH}S. W. Hawking and I. G. Moss, Phys. Lett. {\bf 110B} (1982)
  35. 





\end{thebibliography}
\end{document}